\documentstyle[aps,epsfig,twocolumn]{revtex}
\begin{document}
\draft
\flushbottom
\twocolumn[
\hsize\textwidth\columnwidth\hsize\csname @twocolumnfalse\endcsname

\title{Phase diagram of diluted magnetic semiconductor quantum wells}
\author{L. Brey and F. Guinea}
\address{
Instituto de Ciencia de Materiales (CSIC). Cantoblanco,
28049 Madrid. Spain.} 
\date{\today}
\maketitle 
\tightenlines
\widetext
\advance\leftskip by 57pt
\advance\rightskip by 57pt

\begin{abstract}
The phase diagram of diluted magnetic semiconductor quantum wells is
investigated. The interaction between the carriers 
in the hole gas can
lead to first order ferromagnetic transitions, which remain
abrupt in applied fields. These transitions can be induced
by magnetic fields or, in double-layer  systems by  electric fields. 
We make a number of precise experimental predictions for observing
these first order phase transitions.
\end{abstract}

\pacs{}
]
\narrowtext
\tightenlines

Semiconductor and ferromagnetic materials have complementary properties
in information processing and storage technologies.\cite{prinz}
Recent advances in the MBE growing techniques have made possible the fabrication of 
$Mn$-based diluted magnetic semiconductors (DMS) with a rather high
ferromagnetic-paramagnetic critical temperature $T_c$.\cite{O99}
In semiconductors it is possible to modulate spatially the density of carriers 
by changing
the doping profiles but in DMS it is also suitable  to vary the magnetic 
order  of the carriers by changing the magnetic ion densities. 
The combination of these two possibilities 
opens 
a rich field of applications for these materials.

The high critical temperature DMS's have a high concentration, $c$,
of $Mn ^{2+}$ ions,
randomly located. The itinerant carriers in the $Ga_{1-x}Mn_xAs$
systems are holes and their density, $c ^ *$, is much smaller than
the magnetic ion density.
In doped semiconductor 
the  spin $S$=$5/2$ $Mn^{2+}$ ions feel 
a long range ferromagnetic interaction created by the coupling mediated by 
the itinerant spin polarized carriers.\cite{dietl,taka,Jetal99,LJM00}
This interaction has some
resemblance with the one observed in the 
pyrochlores\cite{Setal96,Aetal99}, where
a similar type of coupling between a narrow electronic band and
magnetic ions is assumed to exist. In the latter case, the phase diagram
is significantly different from that of a conventional 
itinerant ferromagnet, showing first order transitions
and phase separation\cite{GGA00} and/or the formation of localized textures
near the transition temperature\cite{ML98}. These features lead
to interesting transport properties, like colossal magnetoresistance.
We expect these effects to be amplified in low density two-dimensional (2D) 
doped semiconductors, where the carrier-carrier  interaction can play
a significant role, and also favors ferromagnetism\cite{ZOS97}.

In this work, we investigate the nature of the phase diagram
of diluted magnetic semiconductor quantum wells, with emphasis
on the existence of abrupt transitions for experimentally
accessible parameters. 
We analyze quantum wells made of $GaMnAs$ growth in the $< \! 001 \! >$ direction
and with  thickness $w$.
The confinement of the carriers in the $z$-direction can be obtained by
sandwiching the $GaMnAs$ system between non-magnetic $GaAlAs$ semiconductors. 

The system is described by the following Hamiltonian,
\begin{equation}
{\cal H} = {\cal H} _{M}+ {\cal H}_{holes} + J \sum_{I} {\bf S}_I \cdot
{\bf s}_i \, \delta \left( {\bf r}_i - 
{\bf R}_I \right) \, \, \,  .
\label{hamil}
\end{equation}
We now analyze the three terms contributing to  ${\cal H}$.
\par \noindent 
i)${\cal H} _{M}$  is the Hamiltonian of the localized spins interacting
with an external magnetic field $B$. Direct interactions 
between  the magnetic moments of the $Mn$ ions are much smaller than the
interaction with the carrier spins and therefore, unlike in the case
of the pyrochlores,  we
neglect this coupling.
\par \noindent
ii)${\cal H}_{holes}$ is the part of the Hamiltonian which describe the
itinerant holes. It is the sum of the kinetic energy of the holes and the
hole-hole interaction energy.
We treat the kinetic energy of the carriers in the framework of the envelope 
function approximation. The confinement of the carriers in the quantum well 
produces the quantization of their motion 
and the appearance of subbands which,  for 
sufficiently narrow quantum wells,  have a clear light-hole or
heavy-hole character. 
We assume that the hole density is low enough so that only one of the subbands, 
heavy-like, is occupied.
With this, the in plane motion of the holes 
can be approximated by a single parabolic band
of effective mass $m^*$.
We describe the interaction between the electrical carriers with 
the local spin density  approximation (LSDA)\cite{vosko}. Except at very low
densities, the LSDA
gives an accurate description of the electron gas confined in quantum 
wells\cite{EPW92,Ietal99}.
Since the hole $g$-factor is much smaller than the $Mn$ $g$-factor we neglect
in  ${\cal H}_{holes}$ the coupling between the hole spins and the applied magnetic 
field.
\par \noindent
iii)The last term  is the antiferromagnetic exchange interaction  between the
spin of the $Mn^{2+}$ ions located at ${\bf R} _I$ 
and the spins , ${\bf \vec{s}}_i$,
of the itinerant carriers.  The interaction between ions mediated by the
conduction holes is of long range. Thus we will assume
that thermal distribution of the orientation
of the $Mn$ spins is that induced by an effective field, due to the holes,
which should be calculated selfconsistently.

In order to obtain the phase diagram of the DMS system we have to minimize the 
free energy ${\cal F}$. The critical temperatures for the ferromagnetic to
paramagnetic transition in DMS is typically smaller than 100$K$, and for
these temperatures we can consider that the
electron gas is degenerate. Hence, the only temperature
dependence 
in ${\cal F}$ is due to thermal fluctuations of the 
$Mn$ spins. Treating the holes in the mean field approximation the 
free energy per unit area takes the form:
\begin{equation}
{\cal F} = {\cal F}_{ions} + {{ \hbar ^2} \over { m ^*}} {{\pi} \over 2}      
n_{2D} ^2 \, ( 1 + {\xi} ^2 ) + E_{xc} (n_{2D},\xi ) \, \, \, .
\label{freeenergy}
\end{equation}
Here $\xi$ is the carrier spin polarization, $n_{2D}$=$c^* \, w$ is the
two dimensional  density of holes in the system, $E_{xc}$ is the exchange
correlation energy for the holes  and 
${\cal F} _{ions}$ is the contribution of the magnetic ions to the free
energy:
\begin{equation}
{\cal F}_ {ions} = -  T c w  
\log {
{{ \sinh { [ \beta h ( S+ 1/2)]} } \over
{ \sinh { ( \beta h  / 2 )} }  } 
} \, \, \, 
\label{feiones}
\end{equation}
being
\begin{equation}
h= {J \over 2} {{ n_{2d}} \over w} \xi + B \, \, \,
\label{heff}
\end{equation}
the effective magnetic field than the $Mn$ spins feel.
In obtaining Eq.[\ref{feiones}] we have assumed that the 
hole wave function in the $z$-direction has the form $w ^{ -1/2}$.

\begin{figure}
\centerline{\epsfig{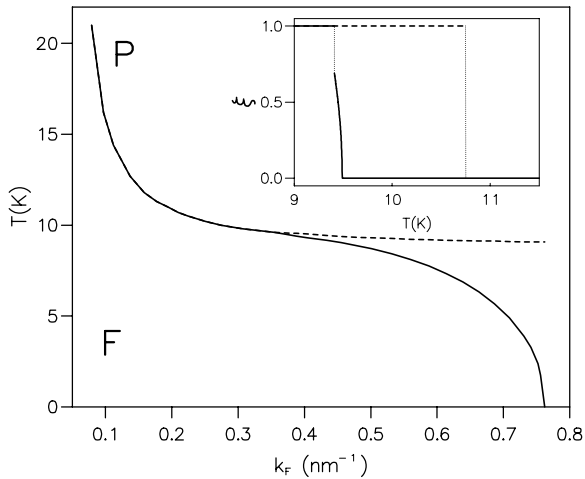}}
\caption{Phase diagram of a magnetic semiconductor quantum well,
using the parameters described in the text. Full line: 
first order transition. Broken line: second order phase transition.
The holes in that phase to the left of the full line are 
completely polarized. The inset shows the polarization of
the holes for $k_F = 0.2$nm$^{-1}$ (broken line) and $k_F =
0.39$nm$^{-1}$.} 
\label{fig:phased}
\end{figure}

The phase diagram with parameters $J = 0.15$eV nm$^{3}$, ion
concentration $c = 1$nm$^{-3}$ and width of the well, $d = 10$nm,
is shown in fig.\{\ref{fig:phased}\}. We include a single hole
band of effective mass $m_{\parallel} = 0.11 m_e$, and 
a dielectric constant $\epsilon_0 = 12.2$\cite{LJM00}.
\vspace{-5.5cm}
The calculations have been done by minimizing the value
of the free energy,
Eq.[\ref{freeenergy}]
and using the expression given by
Vosko {\it et al.} for the exchange correlation potential.
In the high density region the line of continuous transitions agrees with that
obtained from the divergences of the magnetic susceptibility\cite{LJM00}.
The main novelty of our calculation is
the identification of a first order transition to a fully
polarized state at intermediate  densities. 
This transition takes place at higher temperatures
than that at which the magnetic susceptibility of the system diverges.
The existence of a first order transition between the paramagnetic and
the ferromagnetic phases implies the existence of a discontinuity
in the chemical potential at $T_c$.  This discontinuity indicates the
existence of a region in the phase diagram where phase separation 
can take place.
At higher densities, when  a continuous transition between
$\xi =0$ and $\xi  \neq 0$ occurs,
we also find a first order phase transition between the partially 
polarized system, $\xi < 1$, and the fully polarized phase,
$\xi $=1. 

In order to understand better the phase diagram it is 
illustrative to derive the existence of these
first order transitions analytically. For doing that we
treat the hole-hole interaction in the  Hartree-Fock (HF) approximation. 
We obtain that the HF results are in rather good
agreement with the obtained using the LSDA.
In the HF approach the hole energy per unit area can be written
as,\cite{gross}
\begin{eqnarray}
& & E_{holes} =  {{\hbar ^ 2 } \over {m ^* }} { \pi \over 2} n_{2d} ^2 ( 1+ \xi  ^2)
\nonumber \\ & &    - 
{3 \over {8 \pi}} \! {{ e ^2} \over {\epsilon_ 0}} 
\left ( { { 3 \pi ^2 } \over w} \right ) ^{1/3} \! \! 
n _{2d} ^{ 4/3} \left [ ( 1+\xi  ) ^{4/3} \! + \! ( 1-\xi ) ^{4/3} \right ]  ,
\label{electrons}
\end{eqnarray}
where the last term is the exchange energy of the holes. The 
free energy of the ions is given by 
Eq.[\ref{feiones}]. Near a paramagnetic-ferromagnetic transition,
the hole spin polarization is small and we can expand the total free energy in terms of
powers of $\xi$: 
\begin{eqnarray}
  {\cal  F  }_{tot}  & \approx  & 
{{\hbar ^2 } \over { m ^* }} {\pi \over 2}  n ^2 _{2d} 
- C_1 { 3  \over 4  } 
{ { n_{2d} ^ {4/3} } \over {w ^{1 /3} } } 
- T c w \log {(2S+1)}
\nonumber \\  &   +   &   
\xi  ^2   \left   [  
{{\hbar ^2 } \over { m ^* }} {\pi \over 2}  n ^2 _{2d} 
- C_1 {1 \over {6 }} 
{ { n_{2d} ^ {4/3} } \over {w ^{1 /3}} } 
- {1 \over { T }} {c \over w} { { S(S+1)} \over {24}} J^2 n_{2d} ^2 
\right ]
\nonumber \\  & +    &   
\xi  ^4   \left  [   
- C _1 {5 \over {324}} 
{ { n_{2d} ^ {4/3} } \over {w ^{1 /3}} } 
+ {1 \over {  T ^3 }} { c \over {w ^3}} {{ J^ 4
n_{2d} ^4 } \over  {16}}  \times \right. \nonumber \\ 
& &  
\, \, \, \, \,  
\, \, \, \, \,  
\, \, \, \, \,  
\, \, \, \, \,  
\, \, \, \, \, 
\left. \left ( {7 \over {360} } \left( S + {1\over 2} \right) ^ 4
- { { S ^2 ( S+1) ^2 } \over {72}} \right )  
\right ]
 + ...
\label{expansion}
\end{eqnarray}
with
$$
C_1= 
\left ( {  3 \over \pi} \right ) ^ {1/3} 
{ { e ^2}  \over {\epsilon _0} }
\, \, .
\nonumber
$$
We expect a paramagnetic-ferromagnetic transition when the 
quadratic term in $\xi$ is zero. This implies a 
critical temperature 
\begin{equation}
T_C =  {1 \over {12w}}  \, 
{{ c J ^2 S (S+1)} \over 
{ \pi {{ \hbar ^2} \over {m ^* }} -
{1 \over {3 \pi}} 
{ { e ^2}  \over {\epsilon _0}} 
\left ( { {3 \pi ^2} \over w} \right ) ^{1/3}  n_{2d}^{-2/3} }}.
\end{equation}
At high densities, this approximation gives $T_C \approx 8.7$K,
in good  agreement with the LSDA calculation, and with 
the RKKY solution\cite{LJM00}.
Note that the only dependence of $T_c$ on $n_{2d}$ is through the hole-hole 
interaction. This is due to the fact that the two-dimensional density 
of states is energy independent. In three dimensional systems
the kinetic energy scales as $n ^{5/3}$ and $T_c$ is proportional to
$n^ {1/3}$ in agreement with the RKKY theory.
The order of the transition can be inferred from the sign of 
the quartic term in 
Eq.[\ref{expansion}].  If the quartic term is positive the transition 
is of second order, while a negative quartic term 
implies the existence of a first order phase transition.
Using the previous expression
for $T_C$, a first order transition takes place if:
\begin{eqnarray}
{ { e ^2}  \over {\epsilon _0}}  {1 \over { \pi}} 
\left ( { {3 \pi ^2} \over w} \right ) ^{1/3}
{5  \over {2916}} 
{ {n _{2d} ^ {-8/3}} \over { \left ( {7 \over {30}} (S + {1 \over 2} ) ^4 
- {{ S ^2 (S+1) ^2} \over 6} \right ) }} \times
\nonumber \\
{ 
{ c  ^2  J ^2 S ^3 (S+1) ^3 } \over
{
\left ( \pi { {\hbar ^2} \over {m ^* }} \right )  ^3
\left ( 1- { {m ^* } \over { \hbar ^2 }} 
{1 \over { 3 \pi ^2}} 
{ { e ^2}  \over {\epsilon _0}}
\left ( { {3 \pi ^2} \over w} \right ) ^{1/3}
n _ {2d} ^{-2/3} \right ) ^3 }
}
 > 1
\end{eqnarray}
which can be solved to give $n_{2d}  ^{first}  \approx
2.1 \times 10^{12} {\rm cm}^{-2}$. This estimate is also
in good agreement with the LSDA results shown in
fig.\{\ref{fig:phased}\}.

Some comments about the LSDA exchange correlation potential
are in order:
\par \noindent
i)In the HF treatment, the existence of a negative quartic term in the
expansion of the exchange energy in powers of $\xi$ is the
source for the appearance of first order phase transitions.
In the LSDA, the intermediate spin polarization correlation
energy is obtained \cite{barth,gunnarson,vosko,perdew}
by assuming that it has the same polarization dependence 
as the exchange energy, so that
the LSDA expression for $E_{xc}$ also leads to a negative
quartic term when expanded
in powers of $\xi$, and a first order transition is expected.
On the other hand, correlation  effects\cite{gunnarson} 
weaken the spin dependence
of the interaction energy as compared with the results in HF, and
we find that the LSDA gives a value for
$n_{2d} ^{first}$ slightly smaller than the observed in the HF approximation. 
Numerical evaluation\cite{gunnarson}  
of the partially spin polarized $E_{xc}$ also
shows a negative quartic term in 
the expansion of $E_{xc}$ versus $\xi$. Hence,
we believe that the existence of a first order phase transition 
is a {\it robust} result, independent of the model used for
describing the interaction between carriers.
Note, finally, that the existence of a first order transition
implies the absence of long range critical fluctuations, lending further 
support to the adequacy of the methods used in this work.
\par \noindent
ii)The interpolation formula used for describing  the $\xi$ dependence of
$E_{xc}$ is not analytic at $\xi$=1, 
Eq.[\ref{electrons}]. 
This implies that there is, for each pair of values
$n_{2d}$ and $T$, a $\xi$ range near $\xi$=1 which cannot be reached by
minimizing the total free energy. When $\partial {\cal F}^2 / \partial \xi ^2$
is smaller than zero, the system prefers to be completely spin polarized. 
More accurate descriptions 
of the LSDA $E_{xc}$\cite{gunnarson} results in an analytic behavior 
at $\xi$=1. 
Therefore we believe that, in the results shown in 
fig.\{\ref{fig:phased}\}, 
the discontinuous  transition
which occurs 
at lower temperatures than the  second order transition $T_c$,
is, probably,  a spurious result due to the use of a HF-like
interpolation for $E_{xc}$.
\par \noindent
We remark again here that the appearance of a first order transition 
when decreasing $n_{2d}$ is a real result, not related
with the non analyticity of the LSDA expression for $E_{xc}$.

Now we analyze the effect in the phase diagram 
of an external magnetic field.
For $B \neq$0, 
the discontinuous transitions
shown in fig.\{\ref{fig:phased}\} are changed into 
transitions between phases with two different polarizations.
General  thermodynamic arguments show that this line of
first order transitions should end in a critical point, in an
analogous way to the liquid-vapor phase diagram. This
critical point belongs to the Ising universality class.
In the present calculations, the first order transition persists
at all fields. This is due to the non analytic of
the exchange energy at $\xi$=1.
We expect that a correct description of $E_{xc}$ would induce the termination
of the line of first order transitions at a critical value of the field.
Finite temperature effects in the
hole gas, not taken into account, also will weaken the $\xi$
dependence of 
the free energy, leading to the existence
of a critical point.
 
\begin{figure}
\centerline{\epsfig{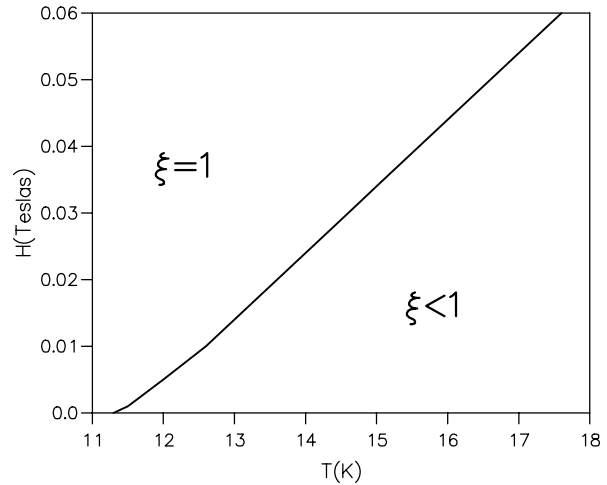}}
\caption{Line of first order phase transitions as
function of applied field and temperature, for fixed density.}
\label{fig:field}
\end{figure}

The line of discontinuous transitions in an applied field
is shown in fig.\{\ref{fig:field}\}.
\vspace{-5.5cm}

The existence of discontinuous transitions leads to
the possibility of phase separation. 
Using Maxwell construction, we find a region of phase
separation near the line of first order transitions. 
In order to analyze its
occurrence, we need to include electrostatic effects.
The simplest situation where phase
separation can be observed is in a bilayer. Let us imagine
two wells with nominal hole density equal to $n_0$. At the
temperatures and fields where the first order transition occurs,
there are two phases with the same free energies, ${\cal F}_1
( n_0 ) = {\cal F}_2 ( n_0 )$. The chemical potential
of these two phases differ at this point, and we define
$\Delta \mu = \mu_1 - \mu_2$. This difference in chemical potentials
will induce a transfer of charge between the two layers, $\delta n$,
until $\Delta \mu$ is compensated by the induced electrostatic
potential, $ V \approx ( e^2 \delta n d ) / 
\epsilon_0$, where $d$ is the distance
between the layers. Thus, we obtain $\delta n \approx ( \Delta \mu
\epsilon_0 ) / ( e^2 d )$. For reasonable values of $d \sim
10 - 50$nm, we find that the charge transfer is small,
$\delta n / n_0 \sim 10^{-2}$. The change in the electrostatic
barrier, $V \approx \Delta \mu$ is, however, comparable 
to the width of the band, and can change significantly
the transport properties.

In a bilayer system, the first order transitions analyzed here
can be induced by an applied electric field. The field induces a difference
in the chemical potential of the two layers, which leads to a
charge transfer. By suitably tuning the parameters, the 
density in one of the layers will reach the value at which
the first order transition discussed above takes place. 
At this point, there will be an abrupt change in the
charge transfer, which can be measured with standard capacitive
techniques\cite{EPW92,Ietal99}. The charge transfer as function of electric
field, for reasonable parameters, is shown in fig.\{{\ref{fig:elec}\}.
 
\vspace{-5.5cm}
\begin{figure}
\centerline{\epsfig{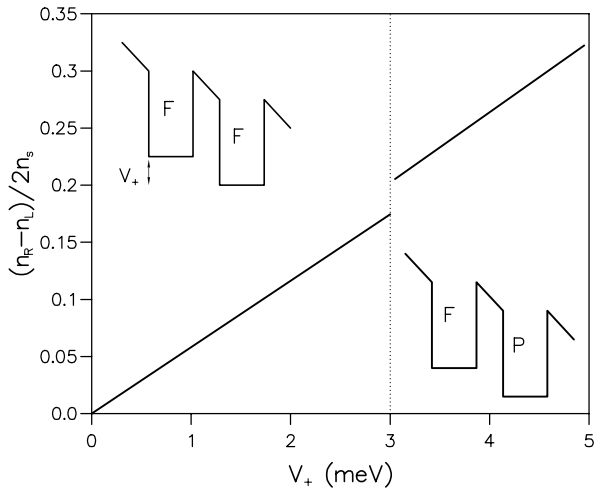}}
\caption{Charge transfer in a bilayer system as function
of the applied electric field.}
\label{fig:elec}
\end{figure}

In conclusion, we have analyzed the possible discontinuous
transitions in two-dimensional  diluted magnetic semiconductors. 
We find that the interaction between carriers  can lead to first order 
transitions in two-dimensional  quantum wells. At the transition, the
holes become fully polarized. This transition can be induced
by a change in the density of the hole gas, the temperature,
a magnetic field, and, in multilayer systems, an applied
electric field. At these transitions, the
minority spin band becomes empty, and the density of states at the Fermi level
is reduced by one half. In double layer  systems, 
significant electrostatic barriers can appear near
the transition. This change can alter significantly
the transport properties\cite{E98}.
Thus, it can be important in the operation
of devices made with these materials. 

LB thanks A.H.MacDonald,  C.Tejedor and A.Rubio  for helpful discussions.
We acknowledge financial support from grants PB96-0875 and PB96-0085 (MEC, Spain)
and (07N/0045/98 and  07N/0027/99) (C. Madrid).

\end{document}